\documentclass[fleqn,12pt]{article}

\usepackage{amssymb}
\usepackage{amsmath}
\usepackage[mathscr]{eucal}
\usepackage{amsthm}
\usepackage{array}
\usepackage{mathbbol}
\usepackage{color}

\swapnumbers
\theoremstyle{definition} 
\newtheorem{defin}{Definition}[section]
\newtheorem{thm}[defin]{Theorem}
\newtheorem{rem}[defin]{Remark}

\newtheorem{cor}[defin]{Corollary}

\newtheorem{lemma}[defin]{Lemma}
\newtheorem{prop}[defin]{Proposition}

\def\vfi{\varphi}
\def\hil{{\mathcal H}}
\def\kil{{\mathcal K}}
\def\A{{\mathcal A}}
\def\B{{\mathcal B}}
\def\C{{\mathcal C}}
\def\F{{\mathcal F}}
\def\half{\frac{1}{2}}

\def\N{\mathbb{N}}
\def\iC{\mathbb{C}}
\def\R{\mathbb{R}}
\def\Z{\mathbb Z}
\def\bz{\left(}
\def\jz{\right)}
\def\bbz{\big(}
\def\bjz{\big)}
\def\inv{^{-1}}
\def\kii{\emph}
\def\kiii{}

\def\egy{\mathbf 1}

\def\dimen{\nu}
\def\torus{\mathbb{T}}
\def\speed{\frac{1}{n^{\dimen}}}
\def\test{\mathcal{T}}
\def\d{\mathrm{d}}

\newcommand{\ki}{\emph}

\newcommand{\s}{\mbox{ }}
\newcommand{\ds}{\mbox{ }\mbox{ }}

\newcommand{\norm}[1]{\left\| #1\right\|}
\newcommand{\snorm}[1]{\| #1 \|}
\newcommand{\inner}[2]{\left\langle #1 , #2\right\rangle}
\newcommand{\binner}[2]{\big\langle #1 , #2\big\rangle}
\newcommand{\abs}[1]{\left| #1 \right|}

\newcommand{\vect}[1]{\mathbf{#1}}

\newcommand{\diad}[2]{|#1\rangle\langle #2|}
\newcommand{\pr}[1]{\diad{#1}{#1}}
\newcommand{\D}{\hat}

\newcommand{\sr}[2]{S(#1\,||\, #2)}
\newcommand{\srm}[2]{S_M(#1\,||\, #2)}

\newcommand{\chernoff}[2]{c(#1\,||\,#2)}
\newcommand{\chernoffli}[2]{\underline c(#1\,||\,#2)}
\newcommand{\chernoffls}[2]{\overline c(#1\,||\,#2)}
\newcommand{\hli}[3]{\underline h(#1|\, #2\,||\,#3)}
\newcommand{\hls}[3]{\overline h(#1|\, #2\,||\,#3)}
\newcommand{\hlim}[3]{h(#1|\, #2\,||\,#3)}
\newcommand{\sli}[2]{\underline s(#1\,||\,#2)}
\newcommand{\sls}[2]{\overline s(#1\,||\,#2)}
\newcommand{\slim}[2]{s(#1\,||\,#2)}
\newcommand{\derleft}[1]{\partial^{-} #1}
\newcommand{\derright}[1]{\partial^{+} #1}
\newcommand{\chbound}[2]{C(#1\,||\,#2)}

\newcommand{\chboundm}[2]{C_M(#1\,||\,#2)}
\newcommand{\hbound}[3]{H(#1|\, #2\,||\,#3)}
\newcommand{\hboundm}[3]{H_M(#1|\, #2\,||\,#3)}

\newcommand{\vecc}[1]{\underline{#1}}

\DeclareMathOperator{\Tr}{Tr}
\DeclareMathOperator{\supp}{supp}

\DeclareMathOperator{\spa}{span}

\DeclareMathOperator{\im}{Im}
\DeclareMathOperator{\re}{Re}
\DeclareMathOperator{\dom}{\mathcal{D}}
\DeclareMathOperator{\ccr}{CCR}
\DeclareMathOperator{\tests}{\mathcal{T}}

\begin{document}

\centerline{\huge Hypothesis testing for Gaussian states on}
\medskip

 \centerline{\huge bosonic lattices}

\bigskip
\s

\bigskip

 \centerline{\large Mil\'an Mosonyi\footnote{Electronic mail: milan.mosonyi@gmail.com}} 
  \bigskip

\centerline{\textit{Graduate School of Information Sciences, Tohoku University}}

\centerline{\textit{Aoba-ku, Sendai 980-8579, Japan}}
\bigskip

 \begin{abstract}
The asymptotic state discrimination problem with simple hypotheses is considered for a cubic lattice of bosons.
A complete solution is provided for 
the problems of the Chernoff and the Hoeffding bounds and Stein's lemma
in the case when both hypotheses are gauge-invariant Gaussian states with translation-invariant quasi-free parts.
\end{abstract}

\section{Introduction}

Assume that we know a priori that the state of an infinite lattice system is either  $\rho_1$
(\ki{null hypothesis} $H_0$) or $\rho_2$ (\ki{alternative hypothesis} $H_1$),
and we want to decide between these two options, based on the outcome of a binary measurement on a finite part of the system. Obviously, there are two ways to make an erroneous decision:
to accept $H_0$ when it is false (\ki{error of the first kind}) and to reject it when it is true (\ki{error of the second kind}). 
In general, one cannot make the corresponding error probabilities to vanish, but they are expected to vanish in the limit as we increase the size of the local system on which the measurement is made.

Hypothesis testing results \cite{Aud,ANSzV,Bjelakovic,Hayashi,HMO,HMO2,HP,HP2,Nagaoka,NSz,ON} show that, in various settings, the optimal error probabilities actually decay exponentially, and the exponent of the optimal decay rate  can be expressed as a certain generalized distance of the states $\rho_1$ and $\rho_2$, depending on the concrete setting of the problem. The most studied cases are the problems of the Chernoff and the Hoeffding bounds and that of Stein's lemma, and the corresponding generalized distances are the Chernoff and the Hoeffding distances and the relative entropy, respectively.
Apart from giving computable closed expressions for the error exponents, the importance of these results lies in providing an operational interpretation for the corresponding generalized distances, which in turn yield alternative and heuristically very transparent proofs for their monotonicity under stochastic operations \cite{Bjelakovic2,Nagaoka}.

The first such result in the quantum setting was 
obtained by Hiai and Petz \cite{HP} and completed later by Ogawa and Nagaoka \cite{ON}, solving the problem of Stein's lemma in an i.i.d.~setting (i.e., when $\rho_1=\bbz\rho_1^{(1)}\bjz^{\otimes\infty}$ and $\rho_2=\bbz\rho_2^{(1)}\bjz^{\otimes\infty}$ are translation-invariant product states) on a one-dimensional spin lattice. This result was later extended 
 to certain correlated situations as well as to higher dimensional lattices \cite{Bjelakovic,HP2}.
The recent findings of Nussbaum and Szko\l a \cite{NSz} and Audenaert et al.~\cite{Aud} created renewed interest in hypothesis testing problems, and their methods were successfully applied to solve such  problems in various settings \cite{Aud,ANSzV,CMMAB,Hayashi,HMO,HMO2,MHOF,Nagaoka,NSz}.

The study of the Chernoff bound for identical copies of one-mode Gaussian states in bosonic systems was initiated in \cite{CMMAB}, where 
an explicit formula for the R\'enyi relative entropies was provided, which was later generalized for $n$-mode states in \cite{PLl}. Stein's lemma for identical copies of one-mode gauge-invariant Gaussian states was treated in \cite{Hayashi2}.
Here we will 
study the hypothesis testing problem for gauge-invariant Gaussian states with translation-invariant quasi-free parts on an infinite bosonic lattice, and we
give a complete solution for the problems of
the Chernoff and the Hoeffding bounds and Stein's lemma in this setting.

The structure of the paper is as follows.
In Section \ref{sec:hypotesting} we give a more technical introduction into hypothesis testing and in Section \ref{sec:CCR} we overview the basic facts about Gaussian states that we will use in the rest of the paper.
In Section \ref{sec:measures} we prove the existence of various asymptotic quantites, including the mean Chernoff and Hoeffding distances and the mean relative entropy, and in Section \ref{sec:exponents} we show that these quantities give the optimal decay rate of the error probabilities in the corresponding settings. In computations with Gaussian states, we will need some basic facts about Fock operators that we collect in a separate Appendix.

\section{Preliminaries}

\subsection{Hypothesis testing on infinite lattice systems}\label{sec:hypotesting}

Consider a lattice system on a $\dimen$-dimensional cubic lattice $\Z^{\dimen}$.
We assume that the observables of the system span a $C^*$-algebra $\A$ and the shift operations on the 
physical lattice lift to an automorphism group $\gamma_{\vect{k}},\,\vect{k}\in\Z^{\dimen}$ on $\A$, 
such that the observable algebra $\A_{\Lambda}$ corresponding to a finite $\Lambda\subset\Z^{\dimen}$ is 
generated by $\{\gamma_{\vect{k}}\bz \A_{\{\vect{0}\}}\jz,\,\vect{k}\in\Lambda\}$. Typical examples are spin 
lattices, when $\A=\otimes_{\vect{k}\in\Z^{\dimen}}\B\bz \iC^d\jz$, and fermionic/bosonic lattices, when 
$\A$ is the CAR/CCR algebra on the anti-symmetric/symmetric Fock space on $l^2\bz\Z^{\dimen}\jz$.
We also assume that for all finite $\Lambda\subset\Z^{\dimen}$ there exists a Hilbert space $\hil_{\Lambda}$ such that 
$\hil_{\Lambda}\cong\otimes_{\vect{k}\in\Lambda}\hil_{\vect{0}}$
and $\A_{\Lambda}\subset\B\bz \hil_{\Lambda}\jz$, which is satisfied by all the examples mentioned above. States of the infinite lattice system are described by positive linear functionals on $\A$ that take the value $1$ on the unit of $\A$.

Assume that we know a priori that the state of the infinite lattice is either $\rho_1$ (null hypothesis $H_0$) or $\rho_2$ (alternative hypothesis $H_1$). To decide between these two hypotheses, we can make 
binary measurements on finite parts of the system, and for simplicity we assume these finite parts to be $\dimen$-dimensional cubes $C_n:=\{\vect{k}\in\Z^{\dimen}\,:\,k_1,\ldots,k_{\dimen}=0,\ldots,n-1\}$. 
A \ki{test} on $C_n$ is an operator $T\in\A_{C_n},\s 0\le T\le I_n$, that determines the binary measurement with measurement operators $T$ and $I-T$. 
If the outcome corresponding to $T$ occurs then $H_0$ is accepted, otherwise it is rejected. 
The error probabilities of the first and the second kind are then expressed as
\begin{equation*}
\alpha_n (T):= \rho_1(I-T)\ds\ds\ds\text{and}\ds\ds\ds
\beta_n (T):=  \rho_2(T).
\end{equation*}

As noted in the Introduction, the error probabilities are expected to decay exponentially if we let $n$ go to infinity and choose the measurements in an optimal way. 
Here we will be interested in the exponents
\begin{eqnarray*}
 \chernoffli{\rho_1}{\rho_2}&:=&\sup_{\{T_n\}}\left\{\liminf_{n\to\infty}-\speed\log\big(
\alpha_n(T_n)+\beta_n(T_n)\big)\right\}\,, \\
 \chernoffls{\rho_1}{\rho_2}&:=&\sup_{\{T_n\}}\left\{\limsup_{n\to\infty}-\speed\log\big(
\alpha_n(T_n)+\beta_n(T_n)\big)\right\}\,, \\
 \chernoff{\rho_1}{\rho_2}&:=&\sup_{\{T_n\}}\left\{\lim_{n\to\infty}-\speed\log\big(
\alpha_n(T_n)+\beta_n(T_n)\big)\right\}\,,
\end{eqnarray*}
corresponding to the problem of the \ki{Chernoff bound},
\begin{eqnarray*}
\hli{r}{\rho_1}{\rho_2}&:=&\sup_{\{T_n\}}\left\{\liminf_{n\to\infty}- \speed\log\beta_n(T_n)\biggm|
\limsup_{n\to\infty} \speed\log\alpha_n(T_n) < -r\right\}\,,\ds r\ge0, \\
\hls{r}{\rho_1}{\rho_2}&:=&\sup_{\{T_n\}}\left\{\limsup_{n\to\infty}- \speed\log\beta_n(T_n)\biggm|
\limsup_{n\to\infty} \speed\log\alpha_n(T_n) < -r\right\}\,,\ds r\ge0, \\
\hlim{r}{\rho_1}{\rho_2}&:=&\sup_{\{T_n\}}\left\{\lim_{n\to\infty} -\speed\log\beta_n(T_n)\biggm|
\limsup_{n\to\infty} \speed\log\alpha_n(T_n) < -r\right\}\,,\ds r\ge0, \label{hb}
\end{eqnarray*}
corresponding to the problem of the \ki{Hoeffding bound}, and
\begin{eqnarray*}
\sli{\rho_1}{\rho_2}&:=&\sup_{\{T_n\}}\left\{\liminf_{n\to\infty}-\speed\log\beta_n(T_n)\biggm| \lim_{n\to\infty}\alpha_n(T_n)=0\right\}\,, \\
\sls{\rho_1}{\rho_2}&:=&\sup_{\{T_n\}}\left\{\limsup_{n\to\infty}-\speed\log\beta_n(T_n)\biggm| \lim_{n\to\infty}\alpha_n(T_n)=0\right\}\,\\
\slim{\rho_1}{\rho_2}&:=&\sup_{\{T_n\}}\left\{ \lim_{n\to\infty}-\speed\log\beta_n(T_n)\biggm| \lim_{n\to\infty}\alpha_n(T_n)=0\right\},
\end{eqnarray*}
corresponding to \ki{Stein's lemma}, respectively. The suprema are taken with respect to sequences of tests, with $T_n\in\A_{C_n},\,n\in\N$.
Obviously,
$\chernoff{\rho_1}{\rho_2}\le\chernoffli{\rho_1}{\rho_2}\le\chernoffls{\rho_1}{\rho_2}$,\\ $\hlim{r}{\rho_1}{\rho_2}\le\hli{r}{\rho_1}{\rho_2}\le\hls{r}{\rho_1}{\rho_2},\, r\ge0$, and $\slim{\rho_1}{\rho_2}\le\sli{\rho_1}{\rho_2}\le\sls{\rho_1}{\rho_2}$.

Assume for the rest that the restrictions $\rho_k^{(n)}$ onto $\A_{C_n}$ 
are given by density operators $\D{\rho}_k^{(n)}$ on $\hil_{C_n}$, i.e., there exist trace-class operators $\D{\rho}_k^{(n)}$ on $\hil_{C_n}$ such that  $\rho_k^{(n)}(a)=\Tr\D{\rho}_k^{(n)}a,\,a\in\A_{\C_n}$. In this case,
\begin{equation*}
\alpha_n (T)=\Tr \D{\rho}_{1}^{(n)} (I_n-T)\ds\ds\ds\text{and}\ds\ds\ds
\beta_n (T)= \Tr \D{\rho}_{2}^{(n)} T.
\end{equation*}
If the supports of $\D{\rho}_1^{(n)}$ and $\D{\rho}_2^{(n)}$ are orthogonal to each other then 
$\rho_1^{(n)}$ and $\rho_2^{(n)}$ can be distinguished perfectly in the sense that there exists a test $T$ for which $\alpha_n(T)=\beta_n(T)=0$, and hence the hypothesis testing problem becomes trivial. To exclude this case, we will assume that 
the supports of $\D{\rho}_1^{(n)}$ and $\D{\rho}_2^{(n)}$ are not orthogonal to each other for any $n\in\N$.
 
Due to H\"older's inequality, the operators $\big(\D{\rho}_{1}^{(n)}\big)^t\big(\D{\rho}_{2}^{(n)}\big)^{1-t}$ are trace-class for each $t\in [0,1]$, and hence the functions
\begin{equation}
\psi_n(t):= \log\Tr \big(\D{\rho}_{1}^{(n)}\big)^t\big(\D{\rho}_{2}^{(n)}\big)^{1-t},\ds\ds\ds t\in [0,1]
\end{equation}
are well-defined for each $n\in\N$, such that 
$-\infty<\psi_n(t)\le 0,\,t\in[0,1],\,n\in\N$. It is 
easy to see that $\psi_n$ is convex on $[0,1]$ for all $n\in\N$.
All along the paper we use the convention $0^t:=0,\,t\in\R$, i.e., we take powers only on the support of $\D{\rho}_{k}^{(n)}$. In particular, $\big(\D{\rho}_{k}^{(n)}\big)^0$ denotes the support projection of $\D{\rho}_k^{(n)}$.
For each $n\in\N$, the \ki{Chernoff disance} of $\rho_1^{(n)}$ and $\rho_2^{(n)}$ 
is defined by 
\begin{equation*}
\chbound{\rho_1^{(n)}}{\rho_2^{(n)}}:=-\inf_{0\le t\le 1}\psi_n(t)\,.
\end{equation*}
For any $r\ge 0$, the \ki{Hoeffding distance} of $\rho_1^{(n)}$ and $\rho_2^{(n)}$ with parameter $r$ is
\begin{equation*}
\hbound{r}{\rho_1^{(n)}}{\rho_2^{(n)}}:=\sup_{0\le t< 1} \frac{-tr-\psi_n(t)}{1-t}.
\end{equation*}

For the rest we assume that the 
additional condition 
%
\begin{equation}\label{assumption}
\bbz\D{\rho}_k^{(n)}\bjz^t,\,k=1,2, \ds \text{are trace-class for all }\ds t\in (0,1]\s\text{ and }\s n\in\N
\end{equation}
holds.
Then, it is not too difficult to see 
(by using the eigen-decompositions and Lebesgue's dominated convergence theorem) 
that $\psi_n$ 
is continuous on $[0,1]$ and differentiable in $(0,1)$ for each $n\in\N$.
Moreover, if  $\supp\D{\rho}_1^{(n)}\le\supp\D{\rho}_2^{(n)}$ then $\psi_n(1)=0$,
and
\begin{equation*}
\hbound{0}{\rho_1^{(n)}}{\rho_2^{(n)}}=\derleft{\psi}_n(1)=\lim_{t\nearrow 1}\psi_n'(t)=
 \Tr\D{\rho}_1^{(n)}\bz\log\D{\rho}_1^{(n)}-\log\D{\rho}_2^{(n)}\jz
=:\sr{\rho_1^{(n)}}{\rho_2^{(n)}},
\end{equation*}
where $\derleft{\psi}_n(1)$ is the left derivative of $\psi_n$ at $1$, and $\sr{\rho_1^{(n)}}{\rho_2^{(n)}}$ is the \ki{relative entropy} of $\rho_1^{(n)}$ and $\rho_2^{(n)}$.
Though assumption \eqref{assumption} is quite restrictive in general, it is automatically satisfied when the local Hilbert spaces are finite-dimensional (which is the case for spin lattices and fermionic lattices) and also when $\rho_1$ and $\rho_2$ are Gaussian states of bosonic lattices, as we will see later.

One can easily see that if the sequence of functions $\frac{1}{n^{\dimen}}\psi_n$ converges uniformly to some function $\psi$ on $[0,1]$ then the \ki{mean Chernoff distance} and the \ki{mean Hoeffding distances} exist, and
\begin{eqnarray}
\chboundm{\rho_1}{\rho_2}&:=&\lim_{n\to\infty}\speed\chbound{\rho_{1}^{(n)}}{\rho_{2}^{(n)}}=-\min_{0\le t\le 1}\psi(t)\,,\label{chboundm}\\
\hboundm{r}{\rho_1}{\rho_2}&:=&\lim_{n\to\infty}\speed\hbound{n^{\dimen}r}{\rho_{1}^{(n)}}{\rho_{2}^{(n)}}=\sup_{0\le t< 1} \frac{-tr-\psi(t)}{1-t}\,,\ds\ds\ds r>0\,.\label{hboundm}
\end{eqnarray}
If, moreover, $\supp\D{\rho}_1^{(n)}\le\supp\D{\rho}_2^{(n)},\,n\in\N$, and  $\lim_n\frac{1}{n^{\dimen}}\derleft{\psi}_n(1)=\derleft{\psi}(1)$ then 
the \ki{mean relative entropy} $\srm{\rho_1}{\rho_2}:=\lim_{n\to\infty}\speed\sr{\rho_1^{(n)}}{\rho_2^{(n)}}$ and the mean Hoeffding distance with parameter $0$ exist, and
\begin{equation}
\srm{\rho_1}{\rho_2}=\hboundm{0}{\rho_1}{\rho_2}
=\derleft{\psi}(1)\,.\label{srm}
\end{equation}

A complete solution to the problems of the Chernoff bound, the Hoeffding bound(s) and to Stein's lemma is obtained if one can show that the relations \eqref{chboundm}, \eqref{hboundm} and \eqref{srm} hold, and
\begin{eqnarray*}
\chernoff{\rho_1}{\rho_2}=\chernoffli{\rho_1}{\rho_2}=\chernoffls{\rho_1}{\rho_2}&=&\chboundm{\rho_1}{\rho_2},\\
\hlim{r}{\rho_1}{\rho_2}=\hli{r}{\rho_1}{\rho_2}=\hls{r}{\rho_1}{\rho_2}&=&\hboundm{r}{\rho_1}{\rho_2},\ds\ds\ds r\ge 0\\
\slim{\rho_1}{\rho_2}=\sli{\rho_1}{\rho_2}=\sls{\rho_1}{\rho_2}&=&\srm{\rho_1}{\rho_2}.
\end{eqnarray*} 
This was done, for instance, for i.i.d.~states on a spin chain \cite{Aud,Hayashi,HP,Nagaoka,NSz,ON}, for quasi-free states on a fermionic lattice \cite{MHOF} and for Gibbs states of translation-invariant finite-range interactions on a spin chain \cite{HMO,HMO2}, apart from the identity $\srm{\rho_1}{\rho_1}=\derleft{\psi}(1)$ (which, however, seems to follow from the results of \cite{Ogata}). Partial results were also obtained for finitely correlated states on spin chains in \cite{HMO,HP2}. Stein's lemma was also proven for 
the case when $\rho_1$ is an ergodic state and $\rho_2$ is a translation-invariant product state on a
 spin lattice \cite{Bjelakovic}.

\subsection{Gaussian states on the CCR algebra}\label{sec:CCR}

Let $(H,\sigma)$ be a \ki{symplectic space}, i.e., $H$ is a real vector space and $\sigma$ is a non-degenerate antisymmetric bilinear form (a \ki{symplectic form}) on $H$, and let $\kappa$ be a positive real number.
We say that a map $W:\,H\to\A$ to a $C^*$-algebra $\A$ is a realization of the $(\kappa,\sigma)$-\ki{canonical commutation relations (CCRs)} if $\A$ is generated by $\{W(x)\,:\,x\in H\}$, and 
\begin{equation*}
W(x)^*=W(-x),\ds\ds
W(x)W(y)=e^{-i\kappa\sigma(x,y)}W(x+y),\ds\ds x,y\in H.
\end{equation*}
Obviously, $\kappa\sigma$ is again a symplectic form, and hence the introduction of $\kappa$ may seem superfluous in the definition. However, we follow this terminology in order to be as compatible as possible with the various conventions appearing in the literature. For a more detailed treatment of the following, we refer the reader to \cite{BR2,Davies,Holevo,Petzccr}.

By Slawny's theorem, any two realizations $W_1:\,H\to\A_1$ and $W_2:\,H\to\A_2$ of the $(\kappa,\sigma)$-CCRs are isomorphic to each other in the sense that there exists a $C^*$-algebra isomorphism $\alpha:\,\A_1\to\A_2$ such that $\alpha\circ W_1=W_2$. 
Moreover, if the symplectic spaces $(H_1,\kappa_1\sigma_1)$ and $(H_2,\kappa_2\sigma_2)$ are isomorphic to each other then any two representations of the corresponding CCRs are also isomorphic to each other. As a consequence, realizations of the $(\kappa,\sigma)$-CCRs with the same $\sigma$ and different $\kappa$'s are all isomorphic to each other, and we denote by $\ccr(H,\sigma)$ the $C^*$-algebra generated by any such realization.
Also, since any two finite-dimensional symplectic spaces of the same dimension are isomorphic to each other, so are the realizations of the corresponding CCRs.
Hence, if $H$ is finite dimensional with $\dim H=2d$ then one can assume without loss of generality that 
$H=\R^{2d}$ and $\sigma$ is its standard symplectic form 
\begin{equation*}
\sigma\bz (x_1,\ldots,x_{2d}),(y_1,\ldots,y_{2d})\jz:=\sum_{k=1}^d\bz x_ky_{k+d}-x_{k+d}y_k\jz,
\end{equation*}
which is the usual choice in physical applications,
and the parameter $\kappa$ is usually taken to be $1/2$ or $1/(2\hbar)$.

Here we will consider the situation when $H=\hil$ for some complex Hilbert space (considered with its real vector space structure) and $\sigma$ is its standard symplectic form $\sigma_{\hil}(x,y):=\im\inner{x}{y},\,x,y\in\hil$.  
Let $\vee^m\hil$ denote the $m$th antisymmetric tensor power of $\hil$, with $\vee^0\hil:=\iC$, and let
$\F(\hil):=\bigoplus_{m=0}^{\infty}\vee^m\hil$
be the \ki{symmetric Fock space}. 
For each $x\in\hil$ let $x_F:=\sum_{m=0}^{\infty}\frac{1}{\sqrt{m!}}x^{\otimes m}\in\F(\hil)$ denote the corresponding \ki{Fock vector} (also called \ki{coherent vector} or \ki{exponential vector}). The Fock vectors are linearly independent and their linear span is dense in $\F(\hil)$.
The \ki{Weyl unitaries} $W_{\kappa}(x),\,x\in\hil$ on $\F(\hil)$ are defined by 
\begin{equation*}
W_{\kappa}(x)y_F:=e^{-\half\kappa\norm{x}^2-\sqrt{\kappa}\inner{x}{y}}(y+\sqrt{\kappa}x)_F,\ds\ds y\in\hil,
\end{equation*}
and they are easily seen to give a realization of the $(\kappa,\sigma_{\hil})$-CCRs. We denote the generated $C^*$-algebra by $\ccr(\hil)$. 

A \ki{state} $\rho$ of $\ccr(\hil)$ is a positive linear functional $\rho:\,\ccr(\hil)\to\iC$, that takes the value $1$ on the unit of $\ccr(\hil)$. 
The \ki{characteristic function} of a state $\rho$ is 
$\hat W_{\kappa}[\rho]:\,H\to\iC,\ds\hat W_{\kappa}[\rho](x):=\rho\bz W_{\kappa}(x)\jz,\,x\in\hil$.
For any real inner product (i.e., positive definite symmetric real bilinear form) $\alpha$ satisfying  
\begin{equation}\label{quasifree1}
\sigma(x,y)^2\le\alpha(x,x)\alpha(y,y),\ds\ds\ds x,y\in\hil,
\end{equation}
there exists a unique state $\rho_{\alpha}$ on $\ccr(\hil)$ with characteristic function
\begin{equation*}
\hat W_{\kappa}[\rho_\alpha](x)=\rho_{\alpha}\bz W_{\kappa}(x)\jz=e^{-\frac{\kappa}{2}\alpha(x,x)},\ds\ds\ds x\in\hil.
\end{equation*}
Such states are called \ki{quasi-free}.
Note that \eqref{quasifree1} is equivalent to the kernel $(x,y)\mapsto \alpha(x,y)+i\sigma(x,y)$ being positive semidefinite.
Obviously, 
\begin{equation*}
\rho_{\alpha,y}(a):=\rho_{\alpha}\bz W_{\kappa}(y)^*aW_{\kappa}(y)\jz,\ds\ds\ds a\in CCR(\hil)
\end{equation*}
is again a state, with characteristic function
\begin{equation*}
\hat W_{\kappa}[\rho_{\alpha,y}](x)=e^{2i\kappa\sigma(y,x)-\frac{\kappa}{2}\alpha(x,x)},\ds\ds\ds x\in\hil.
\end{equation*}
States of this form are called \ki{Gaussian}, and we will refer to $y$ as the \ki{displacement vector}.

The \ki{gauge group} of $\ccr(\hil)$ is the group of quasi-free automorphisms $\gamma_\lambda,\,\lambda\in\torus:=\{z\in\iC\,:\,|z|=1\}$, defined by 
$\gamma_\lambda\bz W_{\kappa}(x)\jz:=W_{\kappa}(\lambda x),\,x\in\hil$.
A state $\rho$ is \ki{gauge-invariant} if $\rho\circ\gamma_\lambda=\rho,\,\lambda\in\torus$.
A Gaussian state $\rho_{\alpha,y}$ is gauge-invariant if and only if $\alpha$ is gauge-invariant, i.e.,
$\alpha(\lambda x,\lambda y)=\alpha(x,y),\,x,y\in\hil,\,\lambda\in\torus$. Moroever, if $\hil$ is finite-dimensional then a Gaussian state $\rho_{\alpha,y}$ is gauge-invariant if and only if there exists a complex linear operator $A\ge I$ such that 
\begin{equation}\label{symbol}
\alpha(x,y)=\re\inner{Ax}{y},\,x,y\in\hil.
\end{equation}
The operator $A$ is called the \ki{symbol} of $\rho_{\alpha,y}$. 
 In the general case we say that $\rho_\alpha$ has a symbol if there exists a complex linear operator $A\ge I$ such that \eqref{symbol} holds. Note that if $\hil$ is infinite-dimensional then having a symbol is a possibly stronger assumption than gauge-invariance.

If $\rho_{\alpha,y}$ is gauge-invariant and $\hil$ is finite dimensional then $\rho_{\alpha,y}$ has a density operator, i.e., there exists a trace-class operator $\D{\rho}_{\alpha,y}$ on $\F(\hil)$ such that $\rho_{\alpha,y}(a)=\Tr\D{\rho}_{\alpha,y} a,\,a\in\ccr(\hil)$. Moreover, the density operator can be expressed in terms of the symbol and the displacement vector in the form
 \begin{equation*}
 \D{\rho}_{\alpha,y}=W_{\kappa}(y)\,\D{\rho}_{\alpha}\,W_{\kappa}(y)^*,\ds\ds
\D{\rho}_{\alpha}=\frac{2^{\dim\hil}}{\det\big( I+A\big)}\bz\frac{A-I}{A+I}\jz_F,
\end{equation*}
where $X_F$ denotes the Fock operator corresponding to an operator $X$ (see \ref{Fock operators}). For a proof, see e.g.~\cite[Corollary 3.2]{Holevo}. Note that the eigenvalues of $A$ coincide with the \ki{symplectic eigenvalues} of $\alpha$, and formula \eqref{normal mode} gives essentially the normal mode decomposition of the state in the above formalism.


Consider now a bosonic lattice system on the $\dimen$-dimensional cubic lattice $\Z^{\dimen}$, such that to each physical site there corresponds one mode of the system. That is, the one-particle Hilbert space of the system is $\hil:=l^2\bz\Z^{\dimen}\jz$
and its observable algebra is $\ccr(\hil)$, the $C^*$-algebra generated by the Weyl unitaries on $\F(\hil)$.
Let 
$\{\egy_{\{\vect{k}\}}\,:\,\vect{k}\in\Z^{\dimen}\}$ denote the standard basis of $l^2\bz\Z^{\dimen}\jz$, let 
$\hil_n:=\spa\{\egy_{\{\vect{k}\}}\,:\,k_1,\ldots,k_{\dimen}=0,\ldots,n-1\}$ and let $P_n$ be the projection onto $\hil_n$. For a bounded operator $A$ on $\hil$, let $A^{(n)}$ denote $P_nAP_n$, when considered as an operator on $\hil_n$.
The Hilbert space of the subsystem corresponding to a cube $C_n:=\{\vect{k}\,:\,k_1,\ldots,k_{\dimen}=0,\ldots,n-1\}$ is $\F(\hil_n)$ and its observable algebra $\A_{C_n}$ is the $C^*$-algebra generated by $\{W_{\kappa}(x)\,:\,x\in\hil_n\}$, that is, $\ccr(\hil_n)$. 

The shift operators are given by $S_{\vect{k}}:\,\egy_{\{\vect{j}\}}\mapsto\egy_{\{\vect{j}+\vect{k}\}},\,\vect{j},\vect{k}\in\Z^{\dimen}$, and they induce the translation automorphisms of $\ccr(\hil)$, given by $\gamma_{\vect{k}}:\,W_{\kappa}(x)\mapsto W_{\kappa}\bz S_{\vect{k}}x\jz,\,x\in\hil,\,\vect{k}\in\Z^{\dimen}$. A state $\rho$ on $\ccr(\hil)$ is \ki{translation-invariant} if $\rho\circ\gamma_{\vect{k}}=\rho,\,\vect{k}\in\Z^{\dimen}$. Assume that $\rho_\alpha$ is a quasi-free state given by a symbol $A$. Then, $\rho_\alpha$ is translation-invariant if and only if $A$ is translation-invariant, i.e., $S_{\vect{k}}AS_{\vect{k}}\inv=A,\,\vect{k}\in\Z^{\dimen}$.

Translation-invariant operators are diagonalized by the Fourier transformation 
\begin{equation*}
\F:\,l^2\bz\Z^{\dimen}\jz\to L^2\bbz[0,2\pi)^{\dimen}\bjz,\ds\ds
\bz\F\egy_{\vect{j}}\jz(\vect{x}):=e^{i\sum_{m=1}^{\dimen}j_mx_m},\ds\ds \vect{x}\in [0,2\pi)^{\dimen},\s\vect{j}\in\Z^{\dimen}.
\end{equation*}
That is, $A$ is translation-invariant if and only if there exists a 
bounded measurable function $a:\,[0,2\pi)^{\dimen}\to \iC$ such that 
$A=\F\inv M_{a}\F$, where $M_{a}$ denotes the multiplication operator by $a$. 
Let $\Sigma(A)$ denote the convex hull of the spectra of $A$.
We will make use of the following multivariate extension of Szeg\H o's theorem \cite{GS}, that was proven in \cite{MHOF}:
\begin{lemma}\label{Szego}
Let $a_1,\ldots,a_r$ be bounded measurable functions on $[0,2\pi)^{\dimen}$ with corresponding shift-invariant operators $A_1,\ldots,A_r$. Then,
\begin{equation}\label{convergence}
\lim_{n\to \infty}\frac{1}{n^{\dimen}}\Tr f_1\bbz A_1^{(n)}\bjz\cdot\ldots\cdot f_r\bbz A_r^{(n)}\bjz=
\frac{1}{(2\pi)^{\dimen}}\int_{[0,2\pi)^{\dimen}} f_1\bz a_1(\vect{x})\jz\cdot\ldots\cdot f_r\bz  a_r(\vect{x})\jz \, d\vect{x}
\end{equation}
for any choice of polynomials $f_1,\ldots,f_r$. If all $a_k$ are real-valued then \eqref{convergence} holds
when $f_k$ is a continuous function on $\Sigma(A_k)$ for all $1\le k\le r$. In this case, the convergence is uniform on norm-bounded subsets of $\prod_{k=1}^n C\bz \Sigma(A_k)\jz$, where 
$C\bz \Sigma(A_k)\jz$ denotes the vector space of continuous functions on $\Sigma(A_k)$, equipped with the supremum norm.
\end{lemma}

\section{Hypothesis testing for Gaussian states}

Consider now the hypothesis testing problem described in Section \ref{sec:hypotesting}. We will assume that 
$\rho_1=\rho_{\alpha_1,y_1}$ and $\rho_2=\rho_{\alpha_2,y_2}$ are both gauge-invariant Gaussian states, and, moreover, that their quasi-free parts $\rho_{\alpha_k}$ are translation-invariant and are given by the symbols $A_k=\F\inv M_{a_k}\F,\,k=1,2$. Here, $a_k:\,[0,2\pi)^{\dimen}\to [1,+\infty)$ are bounded measurable functions.

The restrictions $\rho_{\alpha_k,y_k}^{(n)}$ of $\rho_{\alpha_k,y_k}$ onto $\ccr(\hil_n)$ are again Gaussian states, with symbols
$A_k^{(n)} =P_nA_kP_n$ and displacements $y_k^{(n)}=P_ny_k$.
Moreover, $\rho_{\alpha_k,y_k}^{(n)}$ are given by the density operators $\D{\rho}_{\alpha_k,y_k}^{(n)}=W_{\kappa}(y_k^{(n)})\,\D{\rho}_{\alpha_k}^{(n)}\,W_{\kappa}(y_k^{(n)})^*$,
where $\D{\rho}_{\alpha_k}^{(n)}$ are the densities of the quasi-free parts, and
 \begin{equation*}
 \D{\rho}_{\alpha_k}^{(n)}=\frac{2^n}{\det\big( I+A_k^{(n)}\big)}\bz\frac{A_k^{(n)}-I}{A_k^{(n)}+I}\jz_F=\frac{1}{\det\big( I+ Q_k^{(n)}\big)}\bz\frac{Q_k^{(n)}}{Q_k^{(n)}+I}\jz_F
=
N_{k,n}\bz R_{k,n}\jz_F,
 \end{equation*}
where
\begin{equation}\label{densities}
Q_k:=\bbz A_k-I\bjz/2,\ds\ds R_{k,n}:=\frac{Q_k^{(n)}}{Q_k^{(n)}+I}
\ds\ds\text{and}\ds\ds
N_{k,n}:=1/\det\big( I+ Q_k^{(n)}\big).
\end{equation}
Note that $Q_k=\F\inv M_{q_k}\F$, where $q_k=(a_k-1)/2$ are non-negative bounded measurable functions on $[0,2\pi)^{\dimen}$, and $\rho_{\alpha_k}$ are uniquely determined by either of the functions $a_k,\,q_k$ and
$r_k:=q_k/(1+q_k)$.
 For later use, we define 
\begin{equation*}
W_{n,t}:=R_{n,1}^{t/2}R_{n,2}^{1-t}R_{n,1}^{t/2}\ds\ds\ds\text{and}\ds\ds\ds
w_{t}\bz \vect{x}\jz:=r_1(\vect{x})^{t}
r_2(\vect{x})^{1-t},\ds\ds\ds t\in\R.
\end{equation*}
Note that 
$R_{k,n}\ne R_k^{(n)}:=\bz\F\inv M_{r_k}\F\jz^{(n)}$ and 
$W_{n,t}\ne W_t^{(n)}:=\bz\F\inv M_{w_t}\F\jz^{(n)}$
in general.

 An easy computation yields that for $0<t<1$,
\begin{equation*}
\bbz\D{\rho}_{\alpha_k,y_k}^{(n)}\bjz^t=
M_{k,n,t}\,
\D{\rho}_{f_t (\alpha_k^{(n)}),y},\ds\ds\ds
M_{k,n,t}:=2^{tn}/\det\big[h_{t,-}\big( A_k^{(n)}\big)\big],
\end{equation*}
where 
\begin{equation}\label{ft}
h_{t,\pm}(s):=(s+1)^t\pm (s-1)^t,\ds\ds\ds f_t(s):=\frac{h_{t,+}(s)}{h_{t,-}(s)},\ds\ds\ds t\in(0,1),\s  s\ge 1,
\end{equation}
and
\begin{equation*}
f_t\big(\alpha_k^{(n)}\big)(x,y):=\re\binner{f_t\big( A_k^{(n)}\big) x}{y},\ds\ds\ds x,y\in\hil.
\end{equation*}
That is, the powers $\bbz\D{\rho}_{(\alpha_k,y_k)}^{(n)}\bjz^t$ are again the densities of Gaussian states up to a normalization constant, and hence condition \eqref{assumption} is satisfied.

For some of the statements we will also have to assume that $q_1$ and $q_2$ are strictly positive in the sense that there exists some $\eta>0$ such that $q_k(\vect{x})\ge \eta$ for almost every $\vect{x}$, or equivalently, that $Q_k\ge \eta I$. This assumption ensures that the local restrictions $\rho_{\alpha_k,y_k}^{(n)}$ are faithful for each $n\in\N$, or, in more physical terms, that the vacuum state does not appear in the normal mode decomposition of  $\rho_{\alpha_k}^{(n)}$ for any $n$. Note that this notion of strict positivity is stronger then requiring $q_k(\vect{x})>0$ for almost every $\vect{x}$.

\subsection{Asymptotic distances}\label{sec:measures}

Let $A_k^{(n)}=\sum_j \gamma_{k,n,j}\pr{e_{k,n,j}}$ be eigen-decompositions of the symbols, and assume that the eigenvalues $\gamma_{k,n,j}$ are ordered so that $\gamma_{k,n,j}>1$ for $j\le r(k,n)$ and $\gamma_{k,n,j}=1$ for $j>r(k,n)$. By \eqref{densities}, this gives the eigen-decompositions $R_{k,n}=\sum_{j=1}^{r(k,n)}\tilde\lambda_{k,n,j}\pr{e_{k,n,j}}$, where $\tilde\lambda_{k,n,j}=\frac{\gamma_{k,n,j}-1}{\gamma_{k,n,j}+1}>0,\,j=1,\ldots,r(k,n)$.
With the notations of \eqref{eigenstuff}, we get the eigen-decompositions of the densities as
 \begin{equation*}
 \D{\rho}_{\alpha_k,y}^{(n)}=\sum_{m=0}^{\infty}\,\sum_{m_1+\ldots+m_r=m}\lambda_{k,n,\vecc{m}}\pr{W_{\kappa}(y) e_{k,n,\vecc{m}}},\ds\ds
 \lambda_{k,n,\vecc{m}}:=N_{k,n}\tilde\lambda_{k,n,\vecc{m}}.
 \end{equation*}
Following \cite{NSz}, we define
 \begin{equation*}
 p_{1,n}\bz\vecc{m},\vecc{m}'\jz:=\lambda_{1,n,\vecc{m}}|\inner{e_{1,n,\vecc{m}}}{e_{2,n,\vecc{m}'}}|^2,\ds
p_{2,n}\bz\vecc{m},\vecc{m}'\jz:=\lambda_{2,n,\vecc{m}'}|\inner{e_{1,n,\vecc{m}}}{e_{2,n,\vecc{m}'}}|^2,\ds
 \bz\vecc{m},\vecc{m}'\jz\in J_n,
 \end{equation*}
where 
$J_n:=\{\bz \vecc{m},\vecc{m}'\jz\in \N^{r(1,n)}\times \N^{r(2,n)}\,:\,\inner{e_{1,n,\vecc{m}}}{e_{2,n,\vecc{m}}}\ne 0\}$.
Then, $p_{k,n}$ are positive measures on $J_n$ with $p_{k,n}(J_n)\le 1$, and
\begin{eqnarray*}
 \Tr \bz\D{\rho}_{\alpha_1,y_1}^{(n)}\jz^t\bz\D{\rho}_{\alpha_2,y_2}^{(n)}\jz^{1-t}&=&
\sum_{\bz\vecc{m},\vecc{m}'\jz\in J_n}\lambda_{1,n,\vecc{m}}^t\lambda_{2,n,\vecc{m}'}^{1-t}|\inner{e_{1,n,\vecc{m}}}{e_{2,n,\vecc{m}'}}|^2\\
&=& \sum_{\bz\vecc{m},\vecc{m}'\jz\in J_n}p_{1,n}\bz\vecc{m},\vecc{m}'\jz^tp_{2,n}\bz\vecc{m},\vecc{m}'\jz^{1-t},\ds\ds\ds t\in[0,1].
  \end{eqnarray*}
Define 
\begin{equation*}
\psi_n(t):=\log\sum_{\bz\vecc{m},\vecc{m}'\jz\in J_n}p_{1,n}\bz\vecc{m},\vecc{m}'\jz^tp_{2,n}\bz\vecc{m},\vecc{m}'\jz^{1-t},\ds\ds\ds t\in\R,
\end{equation*}
using the convention $\log +\infty:=+\infty$. By H\"older's inequality, $\psi_n$ is convex on $\R$ for each $n\in\N$.

Our first goal is to express $\psi_n$ in terms of the 
symbols and the displacement vectors. To this end, 
let 
\begin{equation}\label{cnt}
c_{n,t}:=\exp\bbz -2\kappa\big\langle\big[ f_t\bbz A_1^{(n)}\bjz+f_{1-t}\bbz A_2^{(n)}\bjz\big]\inv \overline y,\overline y\big\rangle\bjz,\ds\ds\ds t\in (0,1)
\end{equation}
with $\overline y:=y_2-y_1$ and $f_t$ given in \eqref{ft}.
Note that 
$f_t(s)\ge 1$ for all $t\in(0,1),\,s\ge 1$ and 
$\lim_{t\searrow 0}f_t(s)=+\infty$ if $s> 1$. If $A_1\ne I$ then $A_1^{(n)}\ne I_n,\,n\in\N$, and we have
$
\lim_{t\searrow 0}\big[ f_t\bbz A_1^{(n)}\bjz+f_{1-t}\bbz A_2^{(n)}\bjz\big]\inv=0.
$
If $A_1=I$ (which is the case if and only if $\rho_{\alpha_1}$ is the vacuum state)
then $A_1^{(n)}=I_n,\,n\in\N$, and
$\lim_{t\searrow 0}\left[f_t\bbz A_1^{(n)}\bjz+f_{1-t}\bbz A_2^{(n)}\bjz\right]=I_n+A_2^{(n)}$.
By similar considerations with $A_2$ and $\lim_{t\nearrow 1}$, we get
\begin{equation}\label{cnt1}
c_{n,0}:=\lim_{t\searrow 0}c_{n,t}=
\begin{cases}
1, & \s\text{if} \s A_1\ne I,\\
\exp\bbz -2\kappa\binner{\overline y}{\big[ A_2^{(n)}+I_n\big]\inv\overline y}\bjz, & \s\text{if} \s A_1=I,
\end{cases}
\end{equation}
\begin{equation}\label{cnt2}
c_{n,1}:=\lim_{t\nearrow 1}c_{n,t}=
\begin{cases}
1, & \s\text{if} \s A_2\ne I,\\
\exp\bbz 2\kappa\binner{\overline y}{\big[ A_1^{(n)}+I_n\big]\inv\overline y}\bjz, & \s\text{if} \s A_2=I,
\end{cases}
\end{equation}
for all $n\in\N$.


\begin{lemma}\label{lemma:psi_n}
For each $t\in[0,1]$,
\begin{equation}
\psi_n(t)=
\log c_{n,t}-t\Tr\log\bbz Q_{1}^{(n)}+I_n\bjz-(1-t)\Tr\log\bbz Q_{2}^{(n)}+I_n\bjz-\Tr\log\bbz I_n-W_{n,t}\bjz.\label{formula:psi_n}
 \end{equation}
\end{lemma}
\begin{proof}
The Parseval formula for the Weyl transform \cite{Holevo} tells that for any two Hilbert-Schmidt operators $T_1,T_2$ on $\F(\hil_n)$,
\begin{equation*}
\Tr T_1^*T_2=\bz\frac{\kappa}{\pi}\jz^d\int_{\hil_n}\overline{\Tr (W_{\kappa}(x)T_1)}\Tr( W_{\kappa}(x)T_2)\,d\lambda(x),
\end{equation*}
where $\lambda$ is the Haar-measure on $\hil_n$, normalized so that cubes spanned by symplectic bases have measure $1$.
The choice $T_1:=\bbz\D{\rho}_{(\alpha_1,y_1)}^{(n)}\bjz^t,\,T_2:=\bbz\D{\rho}_{(\alpha_2,y_2)}^{(n)}\bjz^{1-t}$ with $t\in(0,1)$
yields, after some computation,
\begin{eqnarray}
\Tr \bz\D{\rho}_{\alpha_1,y_1}^{(n)}\jz^t\bz\D{\rho}_{\alpha_2,y_2}^{(n)}\jz^{1-t}&=&
c_{n,t}\,\Tr \bz\D{\rho}_{\alpha_1}^{(n)}\jz^t\bz\D{\rho}_{\alpha_2}^{(n)}\jz^{1-t}\nonumber\\
&=&
\frac{2^d c_{n,t}M_{1,n,t}M_{2,n,t}}{\det\big[f_t\bbz A_1^{(n)}\bjz+f_{1-t}\bbz A_2^{(n)}\bjz\big]}\label{PLlbound}\\
&=&
\frac{2^d c_{n,t}}{\det\big[ \bbz A_{1}^{(n)}+I_n\bjz^t\bbz A_{2}^{(n)}+I_n\bjz^{1-t}-\bbz A_{1}^{(n)}-I_n\bjz^t \bbz A_{2}^{(n)}-I_n\bjz^{1-t}\big]}\nonumber\\
&=&
\frac{c_{n,t}}{\det\big[ \bbz Q_{1}^{(n)}+I_n\bjz^t\bbz Q_{2}^{(n)}+I_n\bjz^{1-t}-\bbz Q_{1}^{(n)}\bjz^t\bbz Q_{2}^{(n)}\bjz^{1-t}\big]}.\nonumber
\end{eqnarray}
From the last formula the assertion follows for $t\in (0,1)$, and the cases $t=0$ and $t=1$ 
can be verified by a direct calculation.
\end{proof}
\begin{rem}
Note that the above result gives that for all $t\in[0,1]$,
\begin{equation*}
\psi_n(t)=
\log\Tr \bz\D{\rho}_{\alpha_1,y_1}^{(n)}\jz^t\bz\D{\rho}_{\alpha_2,y_2}^{(n)}\jz^{1-t}
=
\log c_{n,t}+\log\Tr \bz\D{\rho}_{\alpha_1}^{(n)}\jz^t\bz\D{\rho}_{\alpha_2}^{(n)}\jz^{1-t},
\end{equation*}
i.e., the effect of the displacements only appears in the term $\log c_{n,t}$.
\end{rem}

The core of the above Lemma, formula \eqref{PLlbound} was derived in \cite{PLl} in the general (not necessarily gauge-invariant) case. If the two states have the same displacement then the above result might be strengthened and the proof reduces to a straightforward computation,
as shown below:

\begin{lemma}\label{lemma:psi_n2}
Assume that $y_1=y_2=:y$. For all $t\in\R$ such that $W_{n,t}<I_n$,
\begin{equation*}
\psi_n(t)=
-t\Tr\log\bbz Q_{1}^{(n)}+I_n\bjz-(1-t)\Tr\log\bbz Q_{2}^{(n)}+I_n\bjz-\Tr\log\bbz I_n-W_{n,t}\bjz.
\end{equation*}
\end{lemma}
\begin{proof}
With the convention $0^t:=0,\,t\in\R$, the powers $\bbz\D{\rho}_{\alpha_k,y}^{(n)}\bjz^t$ are well-defined positive (not necessarily bounded) operators for any $t\in\R$, and 
\begin{equation*}
\overline{\bbz\D{\rho}_{\alpha_1,y}^{(n)}\bjz^{t/2}
\bbz\D{\rho}_{\alpha_2,y}^{(n)}\bjz^{1-t}
\bbz\D{\rho}_{\alpha_1,y}^{(n)}\bjz^{t/2}}=
N_{1,n}^tN_{2,n}^{1-t}\,W_{\kappa}(y^{(n)})\bz W_{n,t}\jz_F W_{\kappa}(y^{(n)})^*.
\end{equation*}
 By \eqref{trace}, the above operator  
is bounded and trace-class if and only if $W_{n,t}<I$. 

Let us define the trace of a positive operator to be $+\infty$ whenever it is not trace-class. With this convention,
\begin{equation*}
\sum_{\bz\vecc{m},\vecc{m}'\jz\in J_n}\bbz p_{1,n}\bz\vecc{m},\vecc{m}'\jz\bjz^t\bbz p_{2,n}\bz\vecc{m},\vecc{m}'\jz\bjz^{1-t}=
\Tr \overline{\bbz\D{\rho}_{\alpha_1,y}^{(n)}\bjz^{t/2}
\bbz\D{\rho}_{\alpha_2,y}^{(n)}\bjz^{1-t}
\bbz\D{\rho}_{\alpha_1,y}^{(n)}\bjz^{t/2}}
=
N_{1,n}^tN_{2,n}^{1-t}\Tr \bz W_{n,t}\jz_F,
\end{equation*}
from which the assertion follows.
\end{proof}
\begin{rem}
One can easily see that if $q_1$ and $q_2$ are strictly positive then there exists some $\delta>0$ such that $W_{n,t}<I_n$ for all $t\in (-\delta,1+\delta)$. In this sense, Lemma \ref{lemma:psi_n2} is an extension of Lemma \ref{lemma:psi_n} in the case when $y_1=y_2$.
\end{rem}

\medskip

Our next goal is to prove that the limit
\begin{equation}\label{psi}
\psi(t):=\lim_n\frac{1}{n^{\dimen}}\psi_n(t)
\end{equation}
exists for all $t\in[0,1]$. For this, the following simple Lemma will be useful:
\begin{lemma}\label{lemma:bound}
\begin{equation*}\label{bound}
M(q_1,q_2):=\sup\{\norm{W_{n,t}},\norm{w_t}\,:\,t\in[0,1],\,n\in\N\}<1.
\end{equation*}
\end{lemma}
\begin{proof}
Note that $\snorm{Q_k^{(n)}}=\norm{P_n Q_k P_n}\le \norm{Q_k}=\norm{q_k}_{\infty}$, and hence, $\norm{R_{k,n}}\le \norm{q_k}_{\infty}/(1+\norm{q_k}_{\infty})=\norm{r_k}_{\infty}$. Thus for each $t\in[0,1]$,
\begin{equation*}
\norm{W_{n,t}}\le \norm{R_{1,n}}^t\norm{R_{2,n}}^{1-t}\le \norm{r_1}_{\infty}^t\norm{r_2}_{\infty}^{1-t}
\le \max\{\norm{r_1}_{\infty},\norm{r_2}_{\infty}\}<1,
\end{equation*}
and the same bound holds for $\norm{w_t}_{\infty}$, 
from which the assertion follows.
\end{proof}

\begin{lemma}\label{lemma:psi}
The sequence $\frac{1}{n^{\dimen}}\psi_n$ converges uniformly on $[0,1]$ to
\begin{equation}
\psi(t)=-\frac{1}{\bz 2\pi\jz^{\dimen}}\int_{[0,2\pi)^{\dimen}}
\log\left[\bz 1+q_1(\vect{x})\jz^t\bz 1+q_2(\vect{x})\jz^{1-t}-\bz q_1(\vect{x})\jz^t\bz q_2(\vect{x})\jz^{1-t}\right]\,d\vect{x}.\label{psiintegral}
\end{equation}
If $q_1$ and $q_2$ are strictly positive then $\psi$ is twice differentiable in $(0,1)$, and $\psi''(t)>0,\,t\in (0,1)$ unless $q_1(\vect{x})= q_2(\vect{x})$ for almost every $\vect{x}$.
\end{lemma}
\begin{proof}
By lemma \ref{lemma:psi_n},
\begin{equation*}
\psi_n(t)
=\log c_{n,t}-t\Tr\log\big(I_n+Q_1^{(n)}\big)-(1-t)\Tr\log \big(I_n+Q_2^{(n)}\big)-\Tr\log\bz I_n-W_{n,t}\jz.
\end{equation*}
Formulas \eqref{cnt}, \eqref{cnt1} and \eqref{cnt2} show that 
$e^{-\kappa\norm{\overline y}^2}\le c_{n,t}\le 1,\, 0\le t\in 1$,
and hence, 
\begin{equation*}
\lim_n\frac{1}{n^{\dimen}}\log c_{n,t}=0,\ds\ds\ds t\in[0,1].
\end{equation*}

By Szeg\H o's theorem,
\begin{equation*}
\lim_n\frac{1}{n^{\dimen}}\Tr\log\big(I_n+Q_k^{(n)}\big)=\frac{1}{\bz 2\pi\jz^{\dimen}}\int_{[0,2\pi)^{\dimen}}
\log\bz 1+q_k(\vect{x})\jz\,d\vect{x}.
\end{equation*}
By lemma \ref{lemma:bound}, 
$0\le W_{n,t}\le M(q_1,q_2)I_n$ and $0\le w_t(\vect{x})\le M(q_1,q_2)$ for all $t\in [0,1],\,n\in\N$, and almost every $\vect{x}$.
Consider the power series expansion $\log(1-x)=-\sum_{m=0}^{+\infty}\frac{x^m}{m}$, which is absolutely and uniformly convergent on $[0,M(q_1,q_2)]$, and define $p_N(x):=-\sum_{m=0}^N\frac{x^m}{m},\,N\in\N$. Then,
\begin{eqnarray*}
\abs{\frac{1}{n^{\dimen}}\Tr \log\bz I_n-W_{n,t}\jz-\frac{1}{n^{\dimen}}\Tr p_N(W_{n,t})}&\le&
\norm{\log\bz I_n-W_{n,t}\jz- p_N(W_{n,t})}\\
&\le& \max_{x\in [0,M(q_1,q_2)]}\{\abs{\log(1-x)-p_N(x)}\}\xrightarrow[N\to\infty]{}0.
\end{eqnarray*}
Hence, it is enough to show that 
\begin{equation*}
\lim_n\frac{1}{n^{\dimen}}\Tr p_N(W_{n,t})=\frac{1}{\bz 2\pi\jz^{\dimen}}\int_{[0,2\pi)^{\dimen}}
p_N\bz w_{t}\bz \vect{x}\jz\jz\,d\vect{x}
\end{equation*}
for all $N\in\N$.
 This, however, follows immediately from Lemma \ref{Szego}.

The assertion about the differentiability follows from \eqref{psiintegral}, and a straightforward computation yields
\begin{equation*}
\psi''(t)=\frac{1}{2\pi}\int_0^{2\pi}
\bz\frac{\log r_1(\vect{x})-\log r_2(\vect{x})}{1-r_1(\vect{x})^tr_2(\vect{x})^{1-t}}\jz^2
\,d\vect{x},
\end{equation*}
which is strictly positive unless $r_1(\vect{x})=r_2(\vect{x})$ for almost every $\vect{x}$.
\end{proof}

Recall the definitions of the Chernoff and the Hoeffding distances and their mean versions in Section \ref{sec:hypotesting}. The above Lemma yields the following:

\begin{prop}
The mean Chernoff distance exists, and   
\begin{eqnarray*}
\chboundm{\rho_{\alpha_1,y_1}}{\rho_{\alpha_2,y_2}}&=&
-\inf_{0\le t\le 1}\psi(t).
\end{eqnarray*}
If $\supp\D{\rho}_{\alpha_1,y_1}^{(n)}\le \supp\D{\rho}_{\alpha_2,y_2}^{(n)},\,n\in\N$, then the mean Hoeffding distances exist, and 
\begin{eqnarray*}
\hboundm{r}{\rho_{\alpha_1,y_1}}{\rho_{\alpha_2,y_2}}&=&
\sup_{0\le t< 1} \frac{-tr-\psi(t)}{1-t}\,,\ds\ds r>0.
\end{eqnarray*}
\end{prop}
\begin{proof}
The assertions follow immediately from the uniform convergence established in Lemma \ref{lemma:psi}.
\end{proof}

For operators $0<A,B<I$ on a finite-dimensional Hilbert space $\kil$, define
\begin{equation*}
S_2(A\,||\,B):=A\bz\log A-\log B\jz+(I-A)\bz\log\bz I-A\jz-\log\bz I-B\jz\jz,
\end{equation*}
which is a formal generalization of the relative entropy $S_2(a\,||\,b)$ of the Bernoulli distributions $(a,1-a)$ and $(b,1-b)$ that we get when $\dim\kil=1$.

\begin{lemma}\label{lemma:derivatives}
Assume that $q_1$ and $q_2$ are strictly positive. Then,
\begin{eqnarray}
\lim_n\frac{1}{n^{\dimen}}\derleft{\psi}_n(1)&=&
\frac{1}{2\pi}\int_{0}^{2\pi}(1+q_1(\vect{x}))\,S_2(r_1(\vect{x})\,||\,r_2(\vect{x}))\,d\vect{x}=\derleft{\psi}(1)\label{leftder}\\
\lim_n\frac{1}{n^{\dimen}}\derright{\psi}_n(0)&=&
-\frac{1}{2\pi}\int_{0}^{2\pi}(1+q_2(\vect{x}))\,S_2(r_2(\vect{x})\,||\,r_1(\vect{x}))\,d\vect{x}=\derright{\psi}(0).\label{rightder}
\end{eqnarray}
\end{lemma}
\begin{proof}
Define 
$B_n(t):=f_t\bbz A_1^{(n)}\bjz+f_{1-t}\bbz A_2^{(n)}\bjz$ and 
$h_n(t):=\log c_n(t)= -2\kappa\big\langle B_n(t)\inv\overline y,\overline y\rangle$. By a straightforward computation,
\begin{equation*}
h_n'(t)=2\kappa\inner{B_n(t)\inv B_n'(t) B_n(t)\inv\overline y}{\overline y}
=
4\kappa \inner {B_n(t)\inv\left[ X_{1,n,t}\log R_{1,n}-X_{2,n,1-t}\log R_{2,n}\right] B_n(t)\inv\overline y}{\overline y},
\end{equation*}
where 
\begin{equation*}
X_{k,n,t}:=\bbz A_k^{(n)}+I_n\bjz^t\bbz A_k^{(n)}-I_n\bjz^t\big/\left[\bbz A_k^{(n)}+I_n\bjz^t-\bbz A_k^{(n)}-I_n\bjz^t\right]^2,
\end{equation*}
and one can easily see that 
\begin{equation*}
\derleft{h}_n(1)=\lim_{t\nearrow 1}h_n'(t)=-\kappa\inner{\log R_{2,n}\overline y}{\overline y}.
\end{equation*}

A somewhat lengthy but otherwise again straightforward computation yields
\begin{equation*}
\frac{\d}{\d t}\log\det\bz (I_n-W_{n,t}\jz=\Tr\left[\log R_{1,n}-\log R_{2,n}\right]\left[I_n-\bz I_n-W_{n,t}\jz\inv\right],
\end{equation*}
which, in the limit $t\nearrow 1$, yields 
\begin{equation*}
\partial^{-}\log\det\bz I_n-W_{n,t}\jz_{\big|_{t=1}}=\Tr Q_1^{(n)}\left[\log R_{2,n}-\log R_{1,n}\right].
\end{equation*}
Thus, by \eqref{formula:psi_n}, 
\begin{eqnarray}
\derleft{\psi}_n(1)&=&\lim_{t\nearrow 1}\psi_n'(t)\nonumber\\
&=&
-\Tr\log\bbz I_n+Q_1^{(n)}\bjz+\Tr\log\bbz I_n+Q_2^{(n)}\bjz-\Tr Q_1^{(n)}\left[\log R_{2,n}-\log R_{1,n}\right]\nonumber\\
& &
-\kappa\inner{\log R_{2,n}\overline y}{\overline y}\nonumber\\
&=&
\Tr Q_1^{(n)}\left[ \log Q_1^{(n)}-\log Q_2^{(n)}\right]-
\Tr \bbz I_n+Q_1^{(n)}\bjz\left[ \log \bbz I_n+Q_1^{(n)}\bjz-\log \bbz I_n+Q_2^{(n)}\bjz\right]\nonumber\\
& & -\kappa\inner{\log R_{2,n}\overline y}{\overline y}\label{formula2}\\
&=&
\Tr \bbz I_n+Q_1^{(n)}\bjz S_2\bz R_{1,n}\,||\, R_{2,n}\jz
-\kappa\inner{\log R_{2,n}\overline y}{\overline y}.\nonumber
\end{eqnarray}
Since the sequence $\log R_{2,n},\,n\in\N$, is bounded, $\lim_n\frac{1}{n^{\dimen}}\derleft{h}_n(1)=0$, and hence, Lemma \ref{Szego} applied to \eqref{formula2} yields
\begin{eqnarray*}
\lim_n\frac{1}{n^{\dimen}}\derleft{\psi}_n(1)&=&
\frac{1}{2\pi}\int_0^{2\pi}q_1(\vect{x})\left[\log q_1(\vect{x})-\log q_2(\vect{x})\right]\\
& &\ds\ds\ds
-(1+q_1(\vect{x}))\left[\log (1+q_1(\vect{x}))-\log (1+q_2(\vect{x}))\right]\,d\vect{x}\\
&=&
\frac{1}{2\pi}\int_{0}^{2\pi} 
(1+q_1(\vect{x}))S_2(r_1(\vect{x})\,||\,r_2(\vect{x}))\,d\vect{x}\\
&=& \derleft{\psi}(1),
\end{eqnarray*}
where the last identity follows by a straightforward computation from \eqref{psiintegral}. This proves \eqref{leftder}, and \eqref{rightder} follows by a completely similar computation.
\end{proof}


\begin{prop}
Assume that $q_1$ and $q_2$ are strictly positive. Then,
the mean relative entropies $\srm{\rho_{\alpha_1,y_1}}{\rho_{\alpha_2,y_2}}$ and 
$\srm{\rho_{\alpha_2,y_2}}{\rho_{\alpha_1,y_1}}$ and the mean Hoeffding distance
$\hboundm{0}{\rho_{\alpha_1,y_1}}{\rho_{\alpha_2,y_2}}$ exist, and
\begin{eqnarray*}
\srm{\rho_{\alpha_1,y_1}}{\rho_{\alpha_2,y_2}}&=&
\frac{1}{2\pi}\int_{0}^{2\pi}(1+q_1(\vect{x}))\,S_2(r_1(\vect{x})\,||\,r_2(\vect{x}))\,d\vect{x}=\derleft{\psi}(1)=\hboundm{0}{\rho_{\alpha_1,y_1}}{\rho_{\alpha_2,y_2}},\\
\srm{\rho_{\alpha_2,y_2}}{\rho_{\alpha_1,y_1}}&=&
\frac{1}{2\pi}\int_{0}^{2\pi}(1+q_2(\vect{x}))\,S_2(r_2(\vect{x})\,||\,r_1(\vect{x}))\,d\vect{x}=-\derright{\psi}(0).
\end{eqnarray*}
\end{prop}
\begin{proof}
The assertions follow immediately from Lemma \ref{lemma:derivatives} by the identities
$\sr{\rho_{\alpha_1,y_1}^{(n)}}{\rho_{\alpha_2,y_2}^{(n)}}=\derleft{\psi}_n(1)=\hbound{0}{\rho_{\alpha_1,y_1}^{(n)}}{\rho_{\alpha_2,y_2}^{(n)}}$ and 
$\sr{\rho_{\alpha_2,y_2}^{(n)}}{\rho_{\alpha_1,y_1}^{(n)}}=-\derright{\psi}_n(0)$.
\end{proof}

\subsection{Error exponents}\label{sec:exponents}

By the previous section, $\psi(t)=\lim_n\frac{1}{n^{\dimen}}\psi_n(t)$ exists on $[0,1]$. Being the limit of convex functions, $\psi$ is convex as well, and the uniformity of the convergence ensures that $\psi$ is also continuous on $[0,1]$. Moreover, $\psi(t)\le 0,\, t\in [0,1]$. The \ki{polar function} of $\psi$ is 
\begin{equation*}
\vfi(a):=\sup\{ta-\psi(t)\,:\,t\in[0,1]\}.
\end{equation*}

For each $a\in\R$ and $n\in\N$, define the functions
\begin{equation*}
e_{n,a}(T):=e^{-n^{\dimen}a}\alpha_n(T)+\beta_n(T)=
e^{-n^{\dimen}a}-\Tr\left[e^{-n^{\dimen}a}\D{\rho}_{\alpha_1,y}^{(n)}-\D{\rho}_{\alpha_2,y}^{(n)}\right]T
,\ds\ds\ds T\in\test\bz \B\bz \F(\hil_n)\jz\jz.
\end{equation*}
Here we use the notation
\begin{equation*}
\tests(\C):=\{T\in\C\,:\,0\le T\le I\}
\end{equation*}
to denote the set of tests in a $C^*$-algebra $\C$.
Let $S_{n,a}:=\big\{e^{-n^{\dimen}a}\D{\rho}_{\alpha_1,y}^{(n)}-\D{\rho}_{\alpha_2,y}^{(n)}>0\big\}$ be  
the spectral projection corresponding to the positive part of the spectrum of the self-adjoint operator $e^{-n^{\dimen}a}\D{\rho}_{\alpha_1,y}^{(n)}-\D{\rho}_{\alpha_2,y}^{(n)}$. $S_{n,a}$ is usually referred to as the \ki{Neyman-Pearson test} or \ki{Holevo-Helstr\"om test},
and is easily seen to be a minimizer of $e_{n,a}$ on $\test\bz \B\bz \F(\hil_n)\jz\jz$. 

Theorem 1 in \cite{Aud} tells that for positive semidefinite operators $A$ and $B$ on some Hilbert space $\kil$,
\begin{equation*}
\half\Tr(A+B)-\half\Tr|A-B|\le \Tr A^tB^{1-t},\ds\ds\ds t\in [0,1].
\end{equation*}
Substituting $A:=e^{-n^{\dimen}a}\D{\rho}_{\alpha_1,y}^{(n)}$ and $B:=\D{\rho}_{\alpha_2,y}^{(n)}$, we get
\begin{equation*}
e_{n,a}(S_{n,a})\le e^{-tn^{\dimen}a}\Tr\bz\D{\rho}_{\alpha_1,y}^{(n)}\jz^t\bz\D{\rho}_{\alpha_2,y}^{(n)}\jz^{1-t}
,\ds\ds\ds t\in[0,1],
\end{equation*}
and hence,
\begin{equation}\label{Aud2}
\limsup_n\frac{1}{n^{\dimen}}\log e_{n,a}(S_{n,a})\le \inf_{0\le t\le 1}\{-ta+\psi(t)\}=-\vfi(a).
\end{equation}

Note that $S_{n,a}$ is not necessarily in the observable algebra $\ccr(\hil_n)$ of the local system on the cube $C_n$. However, we have the following: 
\begin{lemma}\label{Kaplansky}
For all $a\in\R$, there exists a sequence of tests $\tilde S_{n,a}\in\test(\ccr(\hil_n)),\,n\in\N$ such that 
\begin{eqnarray}
\limsup_{n\to\infty}\frac{1}{n^{\dimen}}e_{n,a}(\tilde S_{n,a})&\le& 
-\vfi(a),\label{pmin}\\
\limsup_{n\rightarrow\infty}\frac{1}{n^{\dimen}}\log\alpha_n(\tilde S_{n,a}) &\le& -\{\vfi(a)-a\},\label{alpha}\\
\limsup_{n\rightarrow\infty}\frac{1}{n^{\dimen}}\log\beta_n(\tilde S_{n,a}) &\le& -\vfi(a)\,.\label{beta}
\end{eqnarray}
\end{lemma}
\begin{proof}
We use that the von Neumann algebra generated by $\ccr(\hil_n)$ is equal to $\B\bz \F(\hil_n)\jz$ \cite[Proposition 5.2.4]{BR2}, and hence it contains $S_{n,a}$ for all $a\in\R$. Then, by Kaplansky's density theorem, there exist tests $\tilde S_{n,a}\in\test(\ccr(\hil_n))$ such that 
\begin{equation*}
e_{n,a}(\tilde S_{n,a})
\le e_{n,a}(S_{n,a})+1/n^{n^{\dimen}}.
\end{equation*}
Hence, \eqref{pmin} follows by \eqref{Aud2}, and the rest is immediate from 
$\beta(\tilde S_{n,a})\le e_{n,a}(\tilde S_{n,a})$ and $e^{-n^{\dimen}a}\alpha_n(\tilde S_{n,a})\le e_{n,a}(\tilde S_{n,a})$.
\end{proof}

For the rest, we will rely on the analysis in the paper \cite{HMO2}. Note that our setting here is somewhat different from that of \cite{HMO2}, as the local algebras are infinite-dimensional and the scaling in the asymptotics is $\frac{1}{n^{\dimen}}$ instead of $\frac{1}{n}$. However, most of the analysis in \cite{HMO2} carries through whenever the existence and differentiability of $\psi$ on $[0,1]$ can be established. Below we show how the results of \cite{HMO2} can be adapted to the present setting.
Note that there are different sign conventions in the literature in defining the error exponents. Here we chose a sign convention opposite to the one in \cite{HMO2} and used a different notation in order to emphasize the equality of the error exponents and the corresponding relative
entropy-like quantities. The correspondence between the notations of this paper and those of \cite{HMO2} is 
\begin{displaymath}
\begin{array}{rclrclrcl}
h\bz r|\, \rho_{\alpha_1,y_1}\,||\,\rho_{\alpha_2,y_2}\jz &=&-B\bz r|\, \vec\rho\,||\,\vec\sigma\jz\,,\ds &
s\bz \rho_{\alpha_1,y_1}\,||\,\rho_{\alpha_2,y_2}\jz&=&-B\bz \vec\rho\,||\,\vec\sigma\jz\,,\\
\underline h\bz r|\, \rho_{\alpha_1,y_1}\,||\,\rho_{\alpha_2,y_2}\jz&=&-\overline B\bz r|\, \vec\rho\,||\,\vec\sigma\jz\,,\ds &
\underline s\bz \rho_{\alpha_1,y_1}\,||\,\rho_{\alpha_2,y_2}\jz&=&-\overline B\bz \vec\rho\,||\,\vec\sigma\jz\,,\\
\overline h\bz r|\, \rho_{\alpha_1,y_1}\,||\,\rho_{\alpha_2,y_2}\jz&=&-\underline B\bz r|\, \vec\rho\,||\,\vec\sigma\jz\,,&
\overline s\bz \rho_{\alpha_1,y_1}\,||\,\rho_{\alpha_2,y_2}\jz&=&-\underline B\bz \vec\rho\,||\,\vec\sigma\jz\,,
\end{array}
\end{displaymath}
with $\dimen=1$ omitted from the notations.

\begin{lemma}\label{lower bound}
Assume that $q_1$ and $q_2$ are strictly positive. For any sequence of tests $T_n\in\tests(\ccr(\hil_n)),\,n\in\N$, and $\derright{\psi}(0)<a<\derleft{\psi}(1)$,
\begin{equation*}
\liminf_{n\to\infty}\frac{1}{n^{\dimen}}e_{n,a}(T_n)\ge 
-\vfi(a).
\end{equation*}
\end{lemma}
\begin{proof}
As $S_{n,a}$ minimizes $e_{n,a}$ over $\tests\bz \B\bz \F(\hil_n)\jz\jz$,
it is enough to show that 
\begin{equation*}
\liminf_{n\to\infty}\frac{1}{n^{\dimen}}e_{n,a}(S_{n,a})\ge 
-\vfi(a).
\end{equation*}
Let
\begin{equation*}
X_{1,n}\bz\vecc{m},\vecc{m}'\jz:=\frac{1}{n^{\dimen}}\log\frac{p_{2,n}\bz\vecc{m},\vecc{m}'\jz}{p_{1,n}\bz\vecc{m},\vecc{m}'\jz}\,,\ds\ds\ds
X_{2,n}\bz\vecc{m},\vecc{m}'\jz:=\frac{1}{n^{\dimen}}\log\frac{p_{1,n}\bz\vecc{m},\vecc{m}'\jz}{p_{2,n}\bz\vecc{m},\vecc{m}'\jz}
\end{equation*}
be random variables on $J_n$. As it was shown in \cite{NSz}, 
\begin{equation}
2e_{n,a}(S_{n,a})\ge
e^{-n^{\dimen}a} p_{1,n}\bz \{X_{1,n}\ge -a\} \jz
 +p_{2,n}\bz \{X_{2,n}>a\}\jz\,,\label{lb1}
\end{equation}
(see also the proof of \cite[Theorem 3.1]{HMO2}). The logarithmic moment generating function of $X_{n,k}$ with respect to $p_{k,n}$ is 
\begin{equation*}
\Phi_{k,n}(t):=\log\int_{\R}e^{tx}\,d \bbz p_{k,n}\circ X_{k,n}^{-1}\bjz=
\log\sum_{\bz\vecc{m},\vecc{m}'\jz\in J_n}e^{tX_{n,k}\bz\vecc{m},\vecc{m}'\jz}\,p_{k,n}\bz\vecc{m},\vecc{m}'\jz,
\end{equation*}
and one can immediately see that
\begin{eqnarray*}
\Phi_{1,n}(n^{\dimen}t)&=&\log\sum_{\bz\vecc{m},\vecc{m}'\jz\in J_n}p_{1,n}\bz\vecc{m},\vecc{m}'\jz^{1-t}p_{2,n}\bz\vecc{m},\vecc{m}'\jz^t =\psi_n(1-t),\\
\Phi_{2,n}(n^{\dimen}t)&=&\log\sum_{\bz\vecc{m},\vecc{m}'\jz\in J_n}p_{1,n}\bz\vecc{m},\vecc{m}'\jz^t p_{2,n}\bz\vecc{m},\vecc{m}'\jz^{1-t}=\psi_n(t).
\end{eqnarray*}
By Lemma \ref{lemma:psi}, the asymptotic logarithmic moment generating functions $\Phi_{k}(t):=\lim_n\frac{1}{n^{\dimen}}\Phi_{k,n}(n^{\dimen}t)$ exist on $[0,1]$ and are differentiable in $(0,1)$, with $\Phi_1(t)=\psi(1-t),\,\Phi_2(t)=\psi(t),\,t\in[0,1]$.
The G\"artner-Ellis theorem (see, e.g.~\cite{DZ}) then yields
\begin{equation*}
\liminf_n\frac{1}{n^{\dimen}}p_{1,n}\bz \{X_{1,n}\ge -a\}\jz\ge -\sup_{0\le t\le 1}\{-at-\Phi_{1}(t)\}
=-\sup_{0\le t\le 1}\{-at-\psi(1-t)\}=a-\vfi(a)
\end{equation*}
and 
\begin{equation*}
\liminf_n\frac{1}{n^{\dimen}}p_{2,n}\bz \{X_{2,n}\ge a\}\jz\ge -\sup_{0\le t\le 1}\{at-\Phi_{2}(t)\}=
-\sup_{0\le t\le 1}\{at-\psi(t)\}=-\vfi(a)
\end{equation*}
for all $\derright{\psi}(0)<a<\derleft{\psi}(1)$. Thus, by \eqref{lb1}, the assertion follows.
\end{proof}

\begin{cor}\label{corollary}
Assume that $q_1$ and $q_2$ are strictly positive. 
For any $\derright{\psi}(0)<a<\derleft{\psi}(1)$, there exists a sequence of tests $\tilde S_{n,a}\in\test(\ccr(\hil_n)),\,n\in\N$, such that 
\begin{eqnarray}
\lim_{n}\frac{1}{n^{\dimen}}e_{n,a}(\tilde S_{n,a})&=&
-\vfi(a),\label{chernoff corollary}\\
\lim_{n}\frac{1}{n^{\dimen}}\log\alpha_n(\tilde S_{n,a}) &=&
 -\{\vfi(a)-a\},\label{alpha2}\\
\lim_n\frac{1}{n^{\dimen}}\log\beta_n(\tilde S_{n,a}) &=& -\vfi(a)\,.\label{beta2}
\end{eqnarray}
\end{cor}
\begin{proof}
The first assertion follows immediately from Lemmas \ref{Kaplansky} and \ref{lower bound}. Lemma 4.4 in 
\cite{HMO2} tells that $\liminf_n\frac{1}{n^{\dimen}}\log\beta_n(T_n)\ge -\vfi(a)$ for any sequence of 
tests for which $\limsup_{n}\frac{1}{n^{\dimen}}\log\alpha_n(T_n)\\\le
 -\{\vfi(a)-a\}$ holds, and hence \eqref{beta2} follows from \eqref{alpha}. Similarly, \eqref{alpha2} 
follows from \eqref{beta} by Remark 4.6 in \cite{HMO2}.
\end{proof}

Now we are in a position to prove our main result:
\begin{thm}
Assume that $q_1$ and $q_2$ are strictly positive and $q_1\ne q_2$. Then,
 \begin{eqnarray}
  \chernoff{\rho_{\alpha_1,y_1}}{\rho_{\alpha_2,y_2}}=\chernoffli{\rho_{\alpha_1,y_1}}{\rho_{\alpha_2,y_2
 }}=\chernoffls{\rho_{\alpha_1,y_1}}{\rho_{\alpha_2,y_2}}&=&\chboundm{\rho_{\alpha_1,y_1}}{\rho_{\alpha_2,y_2}},\ds\ds\ds\label{Chernoff}\\
  \hlim{r}{\rho_{\alpha_1,y_1}}{\rho_{\alpha_2,y_2}}=\hli{r}{\rho_{\alpha_1,y_1}}{\rho_{\alpha_2,y_2}}=\hls{r}{\rho_{\alpha_1,y_1}}{\rho_{\alpha_2,y_2}}&=&\hboundm{r}{\rho_{\alpha_1,y_1}}{\rho_{\alpha_2,y_2}},\ds\ds\ds\label{Hoeffding}\\
  \slim{\rho_{\alpha_1,y_1}}{\rho_{\alpha_2,y_2}}=\sli{\rho_{\alpha_1,y_1}}{\rho_{\alpha_2,y_2}}=\sls{\rho_{\alpha_1,y_1}}{\rho_{\alpha_2,y_2}}&=&\srm{\rho_{\alpha_1,y_1}}{\rho_{\alpha_2,y_2}},\ds\ds\ds\label{Stein}
  \end{eqnarray} 
where \eqref{Hoeffding} holds for all $0\le r< \srm{\rho_{\alpha_2,y_2}}{\rho_{\alpha_1,y_1}}$.
\end{thm}
\begin{proof}
The assumptions yield that $\psi(0)=\psi(1)=1$ and $\derright{\psi}(0)<0<\derleft{\psi}(1)$. Hence, by choosing $a=0$ in Lemma \ref{lower bound}, we get 
\begin{equation*}
\chernoffls{\rho_{\alpha_1,y_1}}{\rho_{\alpha_2,y_2}}\le \vfi(0)=-\min_{0\le t\le 1}\psi(t)=\chboundm{\rho_{\alpha_1,y_1}}{\rho_{\alpha_2,y_2}}.
\end{equation*}
On the other hand, Corollary \ref{corollary} ensures the existence of a sequence of tests $\tilde S_{n,0},\,n\in\N$ such that $\lim_{n}\frac{1}{n^{\dimen}}e_{n,0}(\tilde S_{n,0})=
-\vfi(0)$, and hence, $\chernoff{\rho_{\alpha_1,y_1}}{\rho_{\alpha_2,y_2}}\ge \vfi(0)$, from which \eqref{Chernoff} follows.

To prove \eqref{Hoeffding},
define $\hat \vfi(a):=\vfi(a)-a,\,a\in\R$. By Lemma 4.1 (see also Figure 2) in \cite{HMO2},
$\vfi$ is strictly monotonically increasing on $[\derright{\psi}(0),\derleft{\psi}(1)]$, with range
$[0,\derleft{\psi}(1)]$, while 
$\hat\vfi$ is strictly monotonically decreasing on the same interval, with range
$[0,-\derright{\psi}(0)]$. Hence, for any $0\le r<-\derright{\psi}(0)=\srm{\rho_{\alpha_2,y_2}}{\rho_{\alpha_1,y_1}}$, one can find a unique $a_r\in \left(\derright{\psi}(0),\derleft{\psi}(1)\right]$ such that $\hat\vfi(a_r)=r$. 
By Corollary \ref{corollary}, we have for any 
$\derright{\psi}(0)<a\le a_r$,
\begin{equation*} 
\lim_{n}\frac{1}{n^{\dimen}}\log\alpha_n(\tilde S_{n,a})=
 -\hat\vfi(a)<-\hat\vfi(a_r)=-r \ds\ds\text{and}\ds\ds
\lim_n\frac{1}{n^{\dimen}}\log\beta_n(\tilde S_{n,a})= -\vfi(a).
\end{equation*}
Hence,
\begin{equation*}
 \hlim{r}{\rho_{\alpha_1,y_1}}{\rho_{\alpha_2,y_2}}\ge \sup_{\derright{\psi}(0)<a\le a_r}\vfi(a)=\vfi(a_r),
\end{equation*}
and the rest of the proof 
goes exactly the same way as in the proof of Theorem 4.8 in \cite{HMO2}.

The last assertion follows immediately from Propositions 5.1 and 5.2 in \cite{HMO2}, by noting that in our setting, $\psi(1)=0$ and $\derleft{\psi}(1)=\srm{\rho_{\alpha_1,y_1}}{\rho_{\alpha_2,y_2}}$.
\end{proof}

\section{Conclusion}

We considered the hypothesis testing problem of discriminating two Gaussian states of an infinite bosonic lattice, and gave complete solutions to the problems of the Chernoff bound, the Hoeffding bound and Stein's lemma under the assumptions that both states are gauge-invariant with translation-invariant quasi-free parts. 

Note that the natural structure underlying the theory of the CCR algebra and Gaussian states is a real vector space $H$, equipped with a symplectic form. On the other hand, if $H$ is finite dimensional then a Gaussian state always defines a canonical complexification of $H$ in which the state becomes gauge-invariant. Our assumption that both states are gauge-invariant can heuristically be understood as requiring that the two states yield the same complexification, which is clearly the strongest technical limitation of our approach. It is an open question how to extend our results to the state discrimination problem for non-gauge invariant Gaussian states.

To be able to treat the infinitely extended lattice, we have chosen a $C^*$-algebraic description of the system. In particular, we defined states of the system as linear functionals on the observable algebra, as a density operator may not exist in this case. 
In all computations, however, we used a concrete representation of the CCR algebra, the Fock representation.
Lemma \ref{Kaplansky} shows that, as far as our asymptotic state discrimination problem is concerned, it does not matter 
what representation we use, as the asymptotically optimal performance can be reached by measurement operators from the CCR algebra.

\section*{Acknowledgments}

The author wishes to thank Professors Fumio Hiai, D\'enes Petz and Masahito Hayashi for stimulating discussions on the topic.
Partial funding was provided by the Grant-in-Aid for JSPS
Fellows 18\,$\cdot$\,06916 and the Hungarian Research Grant
OTKA T068258.

\appendix

\def\thesection{Appendix \Alph{section}} 
\section{} \label{Fock operators}

\def\thesection{\Alph{section}} 

If $A\in\B(\hil)$ then $A^{\otimes m}$ leaves $\vee^m\hil$ invariant, and we denote its restriction to $\vee^m\hil$ by $\vee^m A$. The \ki{Fock operator} $A_F$, corresponding to $A$, is
\begin{equation*}
A_F:=\bigoplus_{m=0}^{\infty}\vee^m A\ds\ds\text{with}\ds\ds
\dom(A_F):=\left\{\oplus_{m=0}^{\infty}x_m\in\F(\hil)\,:\,\sum_{m=0}^{\infty}\norm{\bz\vee^m A\jz x_m}^2<\infty\right\}.
\end{equation*}
Note that the Fock operators are closed, and
\begin{equation*}
\F_f(\hil):=\left\{\oplus_{m=0}^M x_m\,:\,x_m\in\vee^m\hil\s 0\le m\le M,\s M\in\N\right\}
\end{equation*}
is a common core for all Fock operators, on which $A_FB_F=(AB)_F$ holds. If $A\ge 0$ then we also have $\bz A_F\jz^t=\bz A^t\jz_F$ on $\F_f(\hil)$ for any $t\in\R$, with the convention $0^t:=0,\,t\in\R$. Fock operators are also characterized by the property $A_Fx_F=\bz Ax\jz_F,\,x\in\hil$.

If $A\ge 0$ is a finite-rank operator and $A=\sum_{k=1}^r \lambda_k\pr{e_k}$ is an eigen-decomposition of $A$, then
\begin{equation}\label{normal mode}
\vee^m A=\sum_{\substack{\vecc{m}\in\N^r \\ m_1+\ldots+m_r=m}}\lambda_{\vecc{m}}\pr{e_{\vecc{m}}}
\end{equation}
is an eigen-decomposition of $\vee^m A$, where 
\begin{equation}\label{eigenstuff}
\lambda_{\vecc{m}}:=\lambda_1^{m_1}\cdot\ldots\cdot\lambda_r^{m_r},\ds\ds\ds
e_{\vecc{m}}:=\frac{1}{\sqrt{m_1!\ldots m_r!m!}}\sum_{\sigma\in S_{m}} U_{\sigma}^{(m)} e_1^{\otimes m_1}\otimes\ldots\otimes e_r^{\otimes m_r},
\end{equation}
and $U_{\sigma}^{(m)},\,\sigma\in S_m$, denotes the standard unitary representation of the symmetric group $S_m$ on $\hil^{\otimes m}$. As a consequence, 
\begin{equation}\label{trace}
\sum_{m=0}^{\infty}\Tr \vee^m A=\prod_{k=1}^r\bz\sum_{m=0}^{\infty}\lambda_k^m\jz,
\end{equation}
which is finite if and only if $A<I$, in which case $A_F$ is trace-class with
\begin{equation}\label{trace formula}
\Tr A_F=\det\bz I-A\jz\inv.
\end{equation}
If $\Gamma(B):=\oplus_{m=0}^{\infty}\Gamma_m(B)$ is the second-quantized version of a finite-rank operator $B\in\B(\hil)$, where $\Gamma_m(B)$ is the restriction of $\sum_{k=1}^m I^{\otimes(k-1)}\otimes B\otimes I^{m-k}$ onto $\vee^m\hil$, then
\begin{equation}\label{trace formula2}
\Tr A_F\Gamma(B)=\frac{1}{\det(I-A)}\Tr\frac{A}{I-A}B\,.
\end{equation}


\end{document}